# Impulse Response Invariant Discretization of Complex Fractional Order Integrator

Zhenlong Wu, YangQuan Chen, *Senior Member IEEE*, and Donghai Li

*Abstract*—When the order of an integrator is a complex number, the integrator is called a complex fractional order integrator (CFOI). The impulse response invariant discretization (IRID) method is proposed to approximately discretize the CFOI. The definition of the CFOI is introduced firstly, and the real and imaginary parts of the CFOI in frequency-domain responses are derived. The code of IRID for the CFOI based on the MATLAB language is explained. The comparisons of the impulse responses and frequency-domain responses between the CFOI and the approximate discrete/continuous transfer functions are presented to illustrate the effectiveness and correctness of the proposed discretization method. This paper offers a reliable method to implement the CFOI.

## I. INTRODUCTION

THIS fractional calculus or non-integer order calculus has attracted wide spread and increasing attention and applications in past decades [1]. The fractional calculus can be used to better characterize many physical phenomena and physical systems such as the movement of calcium sparks [2], capacitor [3] and the pressurized heavy water reactor [4]. In addition, the fractional calculus is also widely used to design the fractional order controllers to enhance the closed-loop system performance such as fractional order proportional integral derivative (FOPID) controller [5], fractional order active disturbance rejection control (FOADRC) [6] and fractional order disturbance observer control (FODOC) [7] etc. where it was shown that fractional order controllers can offer one or more degrees of freedom of the order resulting in the improvement of the control performance. To explore the freedom from using a non-integer order, a natural generalization of fractional order controllers is to consider the complex fractional order controllers where the orders of integrator and differentiator can be any complex value as pioneered in the third generation CRONE controller [8]. The design and tuning of the CFOPID controller has been attempted recently. In [9], the optimization-based tuning methodology is proposed for Complex FOPI (CFOPI) controller. The new structure of the CFOPID controller with the form $PID^{x+jy}$, where $x$ and $y$ are the real and imaginary parts of the order, is designed to ensure the closed-loop system is robust to the gain variations and noises [10]. Besides, the CFOPID controller is designed with the fractional order constraint integral gain optimization (FC-MIGO) algorithm to satisfy the peak value of sensitivity function ($M_s$) and the peak value of the complementary sensitivity function ($M_p$) constraints [11]. The CFOPID controller designed by standardized -chart is applied to a proton exchange membrane fuel cell (PEMFC) system [12].

However, how to reliably digitally implement the CFOPID controller has not been discussed systematically to the authors' best knowledge. In this paper, we focus on fundamental case, the complex fractional order integrator (CFOI), as a special case of the CFOPID controller. The impulse response invariant discretization (IRID) method has been a useful and effective tool to implement the different types of fractional order controllers such as the fractional second order filter with the form $1/(s^2+as+b)^\lambda$ [13], Bode ideal cut-off (BICO) filter with the form $1/\left(\sqrt{1+(s^2/\omega_c^2)}+(s/\omega_c)\right)^{2(1-\eta)}$ [14] and fractional order [proportional derivative] (FO[PD]) with the form $(k_p+k_d s)^\rho$ [15]. In this paper, with the help of the IRID method, the implementation of the CFOI and the pseudo code are presented. The IRID implementation for the CFOI offers a new viewpoint to implement any types of fractional order based on digital impulse responses.

The rest of the paper is organized as follows: the basic mathematical type of the CFOI and its type in frequency-domain are presented in Section 2. In Section 3, the IRID implementation for the CFOI and the pseudo code are discussed. The time and frequency responses of some test codes are shown in Section 4. Finally, Section 5 gives some concluding remarks.

## II. THE DEFINITION OF THE CFOI

The transfer function of the CFOI can be depicted as,

The first author would like to give thanks to the China Scholarship Council (CSC), Grant 201806210219, for funding towards research at University of California, Merced from Sep. 2018 to Sep. 2019. (Corresponding author: YangQuan Chen.)

Z. Wu and D. Li are with the State Key Lab of Power Systems, Department of Energy and Power Engineering, University of Tsinghua, Beijing 100084, China (e-mail: WZLsongshanshan@163.com; lidongh@mail.tsinghua.edu.cn).

Y.Q. Chen is with Mechatronics, Embedded Systems and Automation (MESA) Lab, School of Engineering, University of California, Merced, CA 95343 USA (e-mail: ychen53@ucmerced.edu).

$$G_{CFOI}(s) = \frac{1}{s^{\lambda+j\mu}}, \quad (1)$$

where $s$, $\lambda$ and $\mu$ are the Laplace complex variable, the real and imaginary parts of the CFOI order, respectively. Based on the characteristics of the third generation CRONE controller and the practical realization restrictions in the time domain [16], the structure of the CFOI can be depicted as [11],

$$G_{CFOI}(s) = \left(\text{Re}_{/j}\left(\frac{\omega_{gc}}{s}\right)^{\lambda+j\mu}\right)^{-\text{sign}(\mu)}, \quad (2)$$

where $\text{Re}_{/j}[.]$ is the real part with respect to $j$, $\omega_{gc}$ is the gain-crossover frequency. Equation (2) can be expressed as following,

$$\left(\text{Re}_{/j}\left(\frac{\omega_{gc}}{s}\right)^{\lambda+j\mu}\right)^{-\text{sign}(\mu)} = \left[\left(\frac{\omega_{gc}}{s}\right)^{\lambda}\cos\left(\mu\ln\frac{\omega_{gc}}{s}\right)\right]^{-\text{sign}(\mu)}, \quad (3)$$

where $\lambda$ and $\mu$ are known real parameters of the CFOI, and their values should be located in the ranges of $(0,2)$ and $(-1,0)$, respectively. Then Equation (3) becomes the following equation considering that we have $\mu<1$,

$$\left(\text{Re}_{/j}\left(\frac{\omega_{gc}}{s}\right)^{\lambda+j\mu}\right)^{-\text{sign}(\mu)} = \left(\frac{\omega_{gc}}{s}\right)^{\lambda}\cos\left(\mu\ln\frac{\omega_{gc}}{s}\right), \quad (4)$$

Based on the mathematical identities and the definition in Equation (4), we obtain the frequency-domain response functions of the CFOI as follows,

$$G_{CFOI}(j\omega) = \left(\frac{\omega_{gc}}{\omega}\right)^{\lambda}\frac{\cos\left(\mu\ln\frac{\omega_{gc}}{j\omega}\right)}{j^{\lambda}} = \left(\frac{\omega_{gc}}{\omega}\right)^{\lambda}\frac{A+jB}{C+jD}$$

$$= \left(\frac{\omega_{gc}}{\omega}\right)^{\lambda}\frac{(AC+BD)+j(BC-AD)}{C^2+D^2}, \quad (5)$$

where

$$j^{\lambda} = \cos\left(\frac{\lambda\pi}{2}\right) + j\sin\left(\frac{\lambda\pi}{2}\right) = C + jD \quad (6)$$

and

$$\cos\left(\mu\ln\frac{\omega_{gc}}{j\omega}\right) = \cosh\left(\frac{\mu\pi}{2}\right)\cos\left(\mu\ln\frac{\omega_{gc}}{\omega}\right) + j\sinh\left(\frac{\mu\pi}{2}\right)\sin\left(\mu\ln\frac{\omega_{gc}}{\omega}\right) = A + jB, \quad (7)$$

where, $\cos(jx) = \cosh(x)$, $\sin(jx) = \sinh(x)$ and $C^2 + D^2 = 1$.

The expressions for the real and imaginary parts are as follows,

$$\text{Re}[G_{CFOI}(j\omega)] = \left(\frac{\omega_{gc}}{\omega}\right)^{\lambda}(AC+BD)$$

$$= \cosh\left(\frac{\mu\pi}{2}\right)\cos\left(\mu\ln\frac{\omega_{gc}}{\omega}\right)\cos\left(\frac{\lambda\pi}{2}\right)$$

$$+ \sinh\left(\frac{\mu\pi}{2}\right)\sin\left(\mu\ln\frac{\omega_{gc}}{\omega}\right)\sin\left(\frac{\lambda\pi}{2}\right), \quad (8)$$

and

$$\text{Im}[G_{CFOI}(j\omega)] = \left(\frac{\omega_{gc}}{\omega}\right)^{\lambda}(BC-AD)$$

$$= \sinh\left(\frac{\mu\pi}{2}\right)\sin\left(\mu\ln\frac{\omega_{gc}}{\omega}\right)\cos\left(\frac{\lambda\pi}{2}\right)$$

$$- \cosh\left(\frac{\mu\pi}{2}\right)\cos\left(\mu\ln\frac{\omega_{gc}}{\omega}\right)\sin\left(\frac{\lambda\pi}{2}\right). \quad (9)$$

Equations (8) and (9) for the CFOI can be used to obtain the real frequency-domain response of the CFOI.

## III. THE IRID IMPLEMENTATION FOR THE CFOI

Considering the IRID method has been widely used to implement a fractional second order filter or a fractional integral [13, 17], the IRID method is proposed to implement the CFOI by approximating the impulse response of the CFOI in this section.

The IRID method for the CFOI in (4) can be performed by using the numerical inverse Laplace transform (NILT) technique, which is an algorithm for finding numerical approximation of the inverse Laplace transform for any function defined in "$s$" having the corresponding time-domain impulse response exist.

Based on the data obtained by the IRID method, the approximate discrete/continuous transfer function of the CFOI can be obtained by using Steiglitz-McBride iteration [18]. Then the comparison of the impulse and the frequency-domain responses between the CFOI and the approximate transfer functions can be carried out. The pseudo code is given in Table 1. The implementation code of the CFOI based on the Table 1 can be obtained, which can be accessed in the **supplementary materials section**.

TABLE 1
THE PSEUDO CODE OF THE IMPLEMENTATION OF THE CFOI WITH THE IRID METHOD.

| |
|---|
| *Give the fixed values of $\lambda$, $\mu$ and $\omega_{gc}$;* |
| *Obtain the time-domain impulse response by the MATLAB function nilt; % numerical inverse Laplace transform.* |
| *Obtain the approximate transfer functions of the CFOI by the MATLAB function stmcb; % Steiglitz-McBride iteration.* |
| *Compare the impulse and the frequency-domain responses.* |

## IV. TIME-DOMAIN AND FREQUENCY-DOMAIN RESPONSES OF THE CFOI

Based on Equations (8) and (9), we can obtain the impulse responses of the CFOI with different $\lambda$, $\mu$ and $\omega_{gc}$ as shown in Fig. 1 - Fig. 3. Note that $\mu$ and $\omega_{gc}$ are set as -0.4 and 0.5 in Fig. 1, respectively. $\lambda$ and $\omega_{gc}$ are set as 1.5 and 0.4 in Fig. 2, respectively. $\lambda$ and $\mu$ are set as 1.5 and -0.5 in Fig. 3. From Fig. 1 - Fig. 3, we can observe that different $\lambda$, $\mu$ and $\omega_{gc}$ all have obvious influence on the impulse responses of the CFOI.

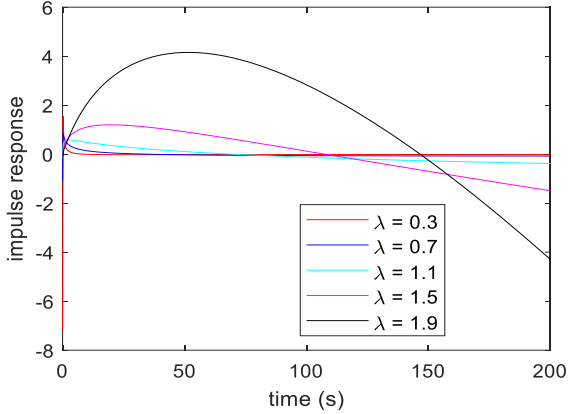

Fig. 1. The impulse responses of the CFOI with different $\lambda$.

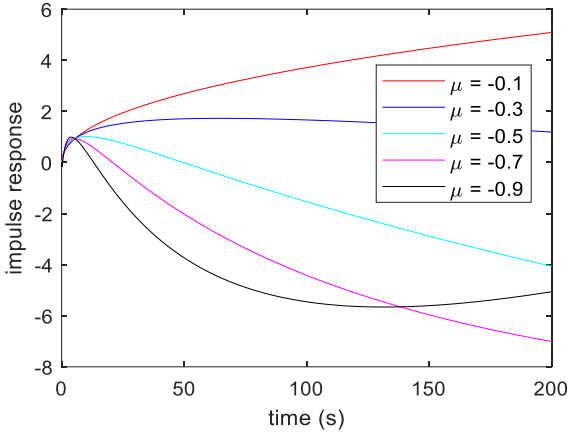

Fig. 2. The impulse responses of the CFOI with different $\mu$.

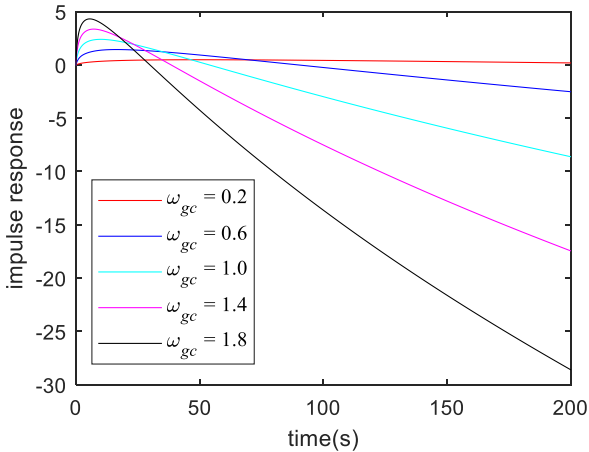

Fig. 3. The impulse responses of the CFOI with different $\omega_{gc}$.

Based on the implementation code of the CFOI based on the Table 1, the approximate discrete/continuous 5$^{th}$ transfer functions of the CFOI are obtained in the form of,

$$G_d\left(z^{-1}\right) = \frac{a_1 z^5 + a_2 z^4 + a_3 z^3 + a_4 z^2 + a_5 z + a_6}{z^5 + b_1 z^4 + b_2 z^3 + b_3 z^2 + b_4 z + b_5}, \quad (10)$$

and

$$G_c(s) = \frac{c_1 s^5 + c_2 s^4 + c_3 s^3 + c_4 s^2 + c_5 s + c_6}{s^5 + d_1 s^4 + d_2 s^3 + d_3 s^2 + d_4 s + d_5}, \quad (11)$$

respectively.

The comparisons of the impulse responses and frequency-domain responses between the CFOI and the approximate discrete/continuous 5$^{th}$ transfer functions are shown in Fig. 4 - Fig. 8. Note that Fig. 5 is the enlarged drawing of Fig. 4. Fig. 4 - Fig. 6 are the results when $\lambda = 1.5$, $\mu = -0.4$ and $\omega_{gc} = 1$. Besides, Fig. 7 - Fig. 8 are the results with $\lambda = 1.5$, $\mu = -0.2$ and $\omega_{gc} = 1$. It can be clearly observe that the impulse responses between the CFOI and the approximate discrete/continuous 5$^{th}$ transfer functions are close even though the frequency-domain responses have some difference in some specific frequency range. Thus this proposed discretization method is "*impulse response invariant*". Besides, the coefficients of the approximate discrete/continuous 5$^{th}$ transfer functions are given in Table 2 and Table 3. The effectiveness of the IRID method for the CFOI can be verified based on the simulations with the function "*[G_opt_d, G_opt_c] = irid_fcoi (lamta, mu, wgc, tm, wmin, wmax, norder)*" as offered in the supplementary materials section.

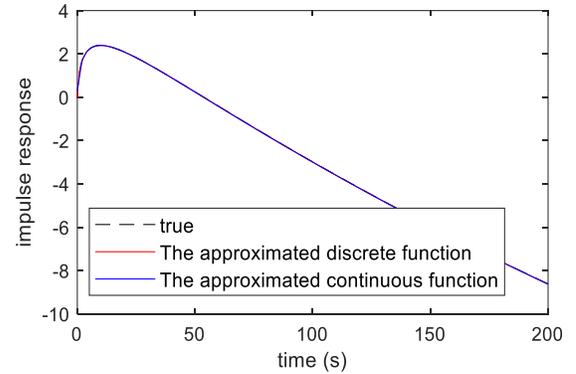

Fig. 4. The comparison in impulse responses with $\lambda = 1.5$, $\mu = -0.4$ and $\omega_{gc} = 1$.

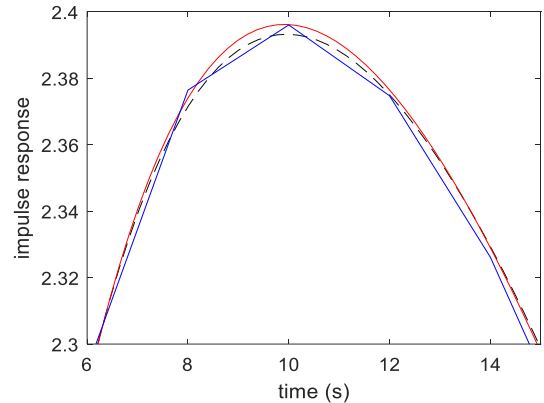

Fig. 5. The enlarged drawing of Fig. 4.

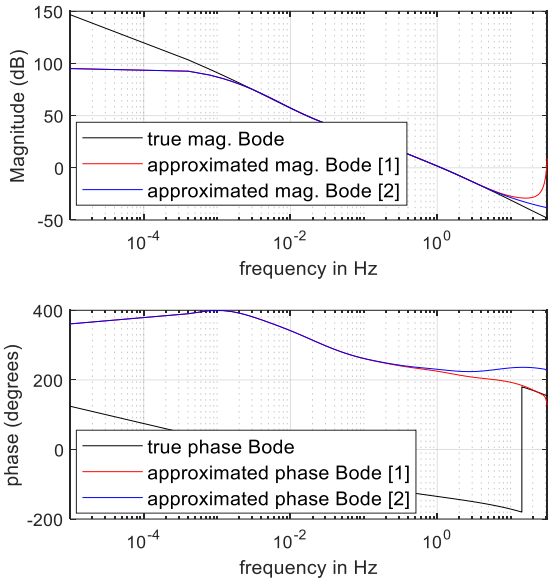

Fig. 6. The comparison in frequency-domain responses with $\lambda = 1.5$, $\mu = -0.4$ and $\omega_{gc} = 1$. ([1]: discrete time, [2]: continuous time.)

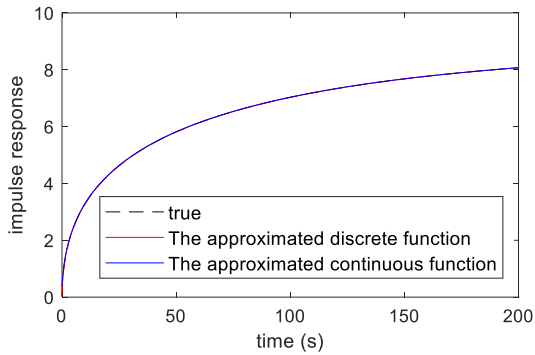

Fig. 7. The comparison in impulse responses with $\lambda = 1.5$, $\mu = -0.2$ and $\omega_{gc} = 1$.

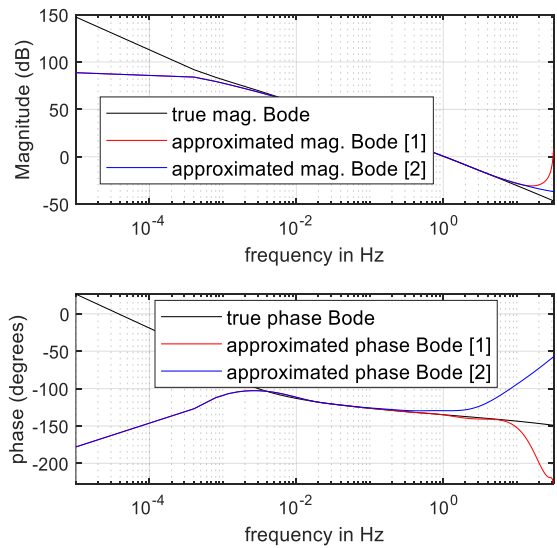

Fig. 8. The comparison in frequency-domain response with $\lambda = 1.5$, $\mu = -0.2$ and $\omega_{gc} = 1$. ([1]: discrete time, [2]: continuous time.)

TABLE 2
THE VALUES OF $a_i / c_i$ ($i = 1, 2, \cdots, 6$)

|  | $\lambda = 1.5$, $\mu = -0.4$ and $\omega_{gc} = 1$ | $\lambda = 1.5$, $\mu = -0.2$ and $\omega_{gc} = 1$ |
| --- | --- | --- |
| $a_1 / c_1$ | -0.0064/-0.006376 | 0.0090/0.008965 |
| $a_2 / c_2$ | 0.1148/0.2791 | 0.0546/0.3991 |
| $a_3 / c_3$ | -0.3195/2.116 | -0.2401/1.864 |
| $a_4 / c_4$ | 0.3416/1.267 | 0.3107/1.102 |
| $a_5 / c_5$ | -0.1518/0.08913 | -0.1639/0.1214 |
| $a_6 / c_6$ | 0.0213/-0.002267 | 0.0298/0.002341 |

TABLE 3
THE VALUES OF $b_i / d_i$ ($i = 1, 2, \cdots, 5$)

|  | $\lambda = 1.5$, $\mu = -0.4$ and $\omega_{gc} = 1$ | $\lambda = 1.5$, $\mu = -0.2$ and $\omega_{gc} = 1$ |
| --- | --- | --- |
| $b_1 / d_1$ | -4.6816/1.829 | -4.7021/1.696 |
| $b_2 / d_2$ | 8.7441/0.5681 | 8.8262/0.4719 |
| $b_3 / d_3$ | -8.1436/0.03439 | -8.2617/0.02781 |
| $b_4 / d_4$ | 3.7803/6.79e-05 | 3.8564/0.0002851 |
| $b_5 / d_5$ | -0.6997/-4.006e-08 | -0.7180/-8.632e-08 |

To further verify the reliability of the proposed method, the impulse responses of the fractional order integrator are compared between the IRID method in this paper and the implementation method in [19]. Note that $\mu$ is set as -0.00001 considering that $\mu$ cannot equal to one as discussed in the definition of the CFOI and the code in [19] can be downloaded and applied from https://www.mathworks.com/matlabcentral/fileexchange/21342-impulse-response-invariant-discretization-of-fractional-order-integrators-differentiators. The impulse responses of $1/s^{0.5}$ and $1/s^{0.8}$ with different methods are presented in Fig. 9 and Fig. 10, respectively. It can be learnt that the impulse responses with different methods are also very close which means that the IRID for the CFOI is reliability and can be used in controller synthesis.

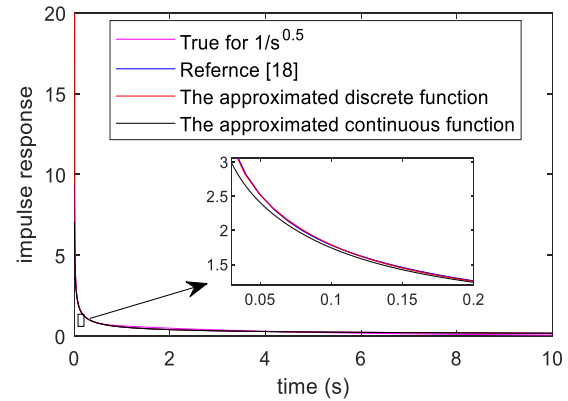

Fig. 9. The impulse responses of $1/s^{0.5}$.

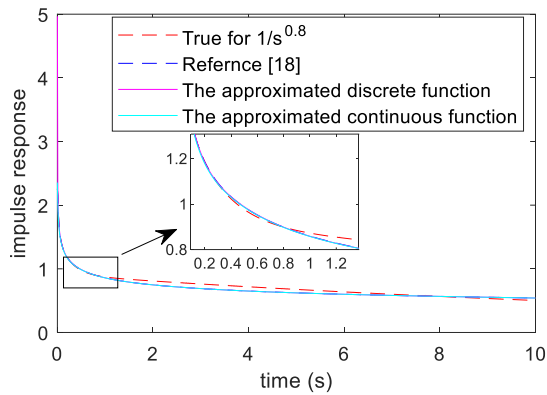

Fig. 10. The impulse responses of $1/s^{0.8}$.

## V. Conclusions

To implement the complex fractional order controller, which allows the orders of integral and derivative actions to be any complex value, this paper proposes the IRID (impulse-response-invariant discretization) method to approximately discretize the CFOI (complex-order fractional integrator). The definition of the CFOI is introduced firstly, and the real and imaginary parts of the CFOI in frequency-domain responses are derived. With the help of the impulse response invariant discretization and the Steiglitz-McBride iteration, the code of IRID for the CFOI based on the MATLAB language is developed. The impulse responses of the CFOI with different $\lambda$, $\mu$ and $\omega_{gc}$, the comparisons of the impulse responses and frequency-domain responses between the CFOI and the approximate discrete/continuous transfer functions are presented in this paper. Besides, the impulse responses of the fractional order integrator are compared by setting the imaginary part of the CFOI as a very small value close to zero. The effectiveness of the IRID method is verified which can ensure the impulse responses between the CFOI and the approximate discrete/continuous transfer functions are very close. This paper offers a reliable method to implement the CFOI and we will extend IRID method to more complex fractional order controllers in the future work.

## VI. Supplementary Materials

This section offers the code to implement the CFOI by the IRID method. Interested readers can download it from https://www.mathworks.com/matlabcentral/fileexchange/73184-irid-of-complex-fractional-order-integrator (MATLAB Central File Exchange) or email to the corresponding author for it.